\documentclass[12pt,twoside,fleqn]{article} 

\usepackage{latexsym} 
\usepackage{amsfonts} 
\usepackage{epsf} 
\usepackage{epsfig}
\usepackage{makeidx} 
\usepackage{graphicx} 

\usepackage{color}

\definecolor{black}{rgb}{0,0,0}
\definecolor{blue}{rgb}{0,0,1}
\definecolor{green}{rgb}{0,1,0}
\definecolor{red}{rgb}{1,0,0}
\definecolor{darkgreen}{rgb}{0,0.7,0}

\makeindex

\newcommand{\ket}[1]{\left|#1\right>} 
\newcommand{\bra}[1]{\left<#1\right|}

\newcommand{\nn}{\nonumber\\} 
 
\newcommand{\f}[1]{\mbox{\boldmath$#1$}}

\newcommand{\na}{\mbox{\boldmath$\nabla$}}
\newcommand{\bea}{\begin{eqnarray}}
\newcommand{\ea}{\end{eqnarray}}
\newcommand{\eea}{\end{eqnarray}}
\newcommand{\ord}{{\cal O}}

 \def\beq{\begin{equation}}
 \def\eeq{\end{equation}}
 \def\bea{\begin{eqnarray}}
 \def\eea{\end{eqnarray}}

\begin{document} 

\title{\bf Cosmological particle creation in the lab?}

\author{Ralf Sch\"utzhold$^{1,*}$ and William G.~Unruh$^{2,+}$
\\
$^1$Fakult\"at f\"ur Physik, Universit\"at Duisburg-Essen,
\\
D-47048 Duisburg, Germany
\\
$^2$Canadian Institute for Advanced Research Cosmology and Gravity Program
\\
Department of Physics and Astronomy, University of British Columbia,
\\
Vancouver B.C., V6T 1Z1 Canada
\\
$^*${\tt ralf.schuetzhold@uni-due.de},
\\
$^+${\tt unruh@physics.ubc.ca}
}

\maketitle

\tableofcontents

\pagestyle{headings}

\newpage

\section{Introduction}

One of the most striking examples for the production of particles out of the 
quantum vacuum due to external conditions is cosmological particle creation,
\index{cosmological particle creation}
which is caused by the expansion or contraction of the Universe.
Already in 1939, Schr\"odinger \index{Schr\"odinger}
understood that the cosmic evolution could lead to a mixing of positive and 
negative frequencies and that this ``would mean production or annihilation of matter, 
merely by the expansion'' [Schr\"odinger, 1939]. 
Later this phenomenon was derived via more modern techniques of quantum 
field theory in curved space-times by Parker [Parker, 1968] \index{Parker}
(who apparently was not aware of Schr\"odinger's work) 
and subsequently has been studied in numerous publications, 
see, e.g., [Birrell \& Davies, 1982; Fulling, 1989; Wald, 1994]. 
Even though cosmological particle creation 
\index{cosmological particle creation}
typically occurs on extremely large length scales, it is one of the very 
few examples for such fundamental effects where we actually may 
have observational evidence:
According to the inflationary model of cosmology, \index{inflation} 
the seeds for the anisotropies 
in the cosmic microwave background (CMB) \index{cosmic microwave background}
and basically all large scale structures stem from this effect, see 
Section~\ref{cosmo-inflation}.
In this Chapter, we shall provide a brief discussion of this phenomenon and 
sketch a possibility for an experimental realization via an analogue in the 
laboratory. \index{laboratory analogues}

\section{Scattering analogy}

For simplicity, let us consider a massive scalar field $\Phi$ 
in the 1+1 dimensional Friedmann-Robertson-Walker metric 
\index{Friedmann-Robertson-Walker metric}
with scale factor $a(\tau)$
\bea\label{cosmol-frw}
ds^2=d\tau^2-a^2(\tau)\,dx^2=a^2(\eta)\left[d\eta^2-dx^2\right] 
\,,
\ea
where $\tau$ is the proper (co-moving) time and $\eta$ the conformal time.
The latter co-ordinate is more convenient for our purpose since the wave 
equation simplifies to 
\bea\label{cosmol-eom}
\left(\frac{\partial^2}{\partial\eta^2}-
\frac{\partial^2}{\partial x^2}-a^2(\eta)\,m^2\right)\Phi(\eta,x)
\,.
\ea
In the massless case $m=0$, the scalar field is conformally invariant 
(in 1+1 dimensions) and thus the expansion does only create particles 
for $m>0$. 
After a spatial Fourier transform, we find that each mode $\phi_k(\eta)$ 
behaves like a harmonic oscillator with a time-dependent potential
\bea\label{cosmol-oscillator}
\left(\frac{d^2}{dt^2}+\Omega^2(t)\right)\phi(t)=0
\,,
\ea
with $k^2+a^2(\eta)\,m^2\to\Omega^2(t)$ and $\eta\to t$. 
There is yet another analogy which might be interesting to notice.
If we compare the above equation to a Schr\"odinger \index{Schr\"odinger}
scattering problem in one spatial dimension
\bea\label{cosmol-Schrodinger}
\left(-\frac{1}{2m}\,\frac{d^2}{dx^2}+V(x)\right)\Psi(x)=E\Psi(x)
\,,
\ea
we find that is has precisely the same form after identifying 
$t\leftrightarrow x$,
$\phi(t)\leftrightarrow\Psi(x)$, and $\Omega^2(t)\leftrightarrow 2m[E-V(x)]$. 
Note that $\Omega^2$ is always greater than zero in our case -- which 
corresponds to propagation over the barrier $E>V(x)$.
If $\Omega^2$ were less than zero over some region in time, 
one would have a barrier penetration (i.e., tunnelling) problem $E<V(x)$. 
With the condition that in the past the field has the form
$e^{i\Omega_{\rm in}t}$, in the future the solution would be 
$\alpha e^{i\Omega_{\rm out}t}+\beta e^{-i\Omega_{\rm out}t}$ 
due to scattering from the region where $\Omega^2<0$. 
This would correspond to particle creation with probability proportional 
to $|\beta|^2$. 
However even if $\Omega^2>0$ everywhere there will still be some 
scattering (above the barrier).

In order to derive the cosmological particle creation, 
\index{cosmological particle creation}
we can study a positive pseudo-norm solution of Eq.~(\ref{cosmol-oscillator}) 
which initially behaves as $e^{-i\Omega_{\rm in} t}$ 
and finally evolves into a 
mixture of positive and negative pseudo-norm solutions -- 
which is in this case equivalent to positive and 
negative frequencies 
$\alpha e^{-i\Omega_{\rm out} t}+\beta e^{+i\Omega_{\rm out} t}$ 
(assuming that $\Omega$ is constant asymptotically). 
In the Schr\"odinger \index{Schr\"odinger}
scattering problem, the initial solution $e^{-i\Omega t}$ 
could be identified with a left-moving wave on the left-hand side of the 
potential ``barrier'' while the final solution 
$\alpha e^{-i\Omega t}+\beta e^{+i\Omega t}$ would then correspond to a 
mixture 
of left-moving $\alpha e^{-i\Omega t}$  and right-moving 
$\beta e^{+i\Omega t}$ waves on the right-hand-side.
As a consequence, the Bogoliubov coefficients $\alpha$  and $\beta$ 
\index{Bogoliubov coefficients} 
are related to the reflection $R$ and transmission $T$ coefficients via 
$\alpha=1/T$  and $\beta=R/T$. 
In this way, the Bogoliubov relation $|\alpha|^2-|\beta|^2=1$ is equivalent 
to the conservation law $|R|^2+|T|^2=1$ for the Schr\"odinger 
\index{Schr\"odinger}
scattering problem.
The probability for particle creation can be inferred from the expectation
value of the number of final particles in the initial vacuum state which 
reads $\bra{0_{\rm in}}\hat n_{\rm out}\ket{0_{\rm in}}=|\beta|^2$. 

\section{WKB analysis}

\index{WKB method}

In order to actually calculate or estimate the Bogoliubov coefficients,
\index{Bogoliubov coefficients} 
let us re-write Eq.~(\ref{cosmol-oscillator}) in a first-order form via 
introducing the phase-space vector $\f{u}$ and the matrix $\f{M}$ 
\bea\label{cosmol-first-order}
\frac{d}{dt}
\left(
\begin{array}{c}
\phi \\
\dot\phi
\end{array}
\right)
=
\f{\dot u}
=
\left(
\begin{array}{cc}
0 & 1 \\
-\Omega^2(t) & 0
\end{array}
\right)
\cdot
\left(
\begin{array}{c}
\phi \\
\dot\phi
\end{array}
\right)
=
\f{M}\cdot\f{u}
\,.
\ea
If we define an inner product via \index{inner product}
\bea\label{cosmol-inner}
(\f{u}|\f{u'})=i(u_2^*u_1'-u_1^*u_2')
\,,
\ea
we find that the inner product of two solutions $\f{u}$ and $\f{u'}$ of 
Eq.~(\ref{cosmol-first-order}) is conserved 
\bea\label{cosmol-conserved}
\frac{d}{dt}(\f{u}|\f{u'})=0
\,.
\ea
The split of a solution into positive and negative frequencies 
(i.e., positive and negative pseudo-norm) corresponds 
to a decomposition in the instantaneous eigen-basis of the matrix 
\bea\label{cosmol-instantaneous}
\f{M}\cdot\f{u}_\pm
=
\pm i\Omega\f{u}_\pm
\,.
\ea
Choosing the usual normalization $\f{u}_\pm=(1,\pm i\Omega)^T/\sqrt{2\Omega}$, 
we find 
\bea\label{cosmol-ortho}
(\f{u}_+|\f{u}_+)=1
\,,\;
(\f{u}_-|\f{u}_-)=-1
\,,\;
(\f{u}_+|\f{u}_-)=0
\,.
\ea
At each time $t$, we may expand a given solution $\f{u}(t)$ of 
Eq.~(\ref{cosmol-first-order}) into the instantaneous eigen-vectors
\bea\label{cosmol-expand}
\f{u}(t)
=
\alpha(t)e^{i\varphi(t)}\f{u}_+(t)
+
\beta(t)e^{-i\varphi(t)}\f{u}_-(t)
\,,
\ea
where the pre-factors are now defined as time-dependent 
Bogoliubov coefficients $\alpha(t)$  and $\beta(t)$.
\index{Bogoliubov coefficients}
It is useful to separate out the oscillatory part with the WKB phase 
\index{WKB phase}
\bea\label{cosmol-phase}
\varphi(t)=\int\limits_{-\infty}^t dt'\,\Omega(t')
\,.
\ea
Now we may insert the expansion (\ref{cosmol-expand}) into the equation of 
motion 
(\ref{cosmol-first-order}) and project it with the inner product
\index{inner product}
(\ref{cosmol-inner}) onto the eigen-vectors $\f{u}_\pm$  which gives 
\bea\label{cosmol-evolution}
\dot\alpha=\frac{\dot\Omega}{2\Omega}\,e^{-2i\varphi}\beta
\,,\;
\dot\beta=\frac{\dot\Omega}{2\Omega}\,e^{2i\varphi}\alpha
\,,
\ea
due to $(\f{u}_-|\f{\dot u}_+)=\dot\Omega/(2\Omega)$ and 
$(\f{u}_+|\f{\dot u}_-)=-\dot\Omega/(2\Omega)$ while 
$(\f{u}_+|\f{\dot u}_+)=(\f{u}_-|\f{\dot u}_-)=0$. 

This equation (\ref{cosmol-evolution}) is still exact and very hard to solve 
analytically -- except in very special cases.
It can be solved formally by a iterative integral equation 
\bea\label{cosmol-iterative}
\alpha_{n+1}
&=&
\alpha_{\rm in}+\int\limits_{-\infty}^t dt'\,
\frac{\dot\Omega(t')}{2\Omega(t')}\,e^{-2i\varphi(t')}\beta_n(t')
\,,
\nn
\beta_{n+1}
&=&
\beta_{\rm in}+\int\limits_{-\infty}^t dt'\,
\frac{\dot\Omega(t')}{2\Omega(t')}\,e^{-2i\varphi(t')}\alpha_n(t')
\,.
\ea
It can be shown that this iteration converges to the exact solution for 
well-behaved $\Omega(t)$ [Braid, 1970].
Standard perturbation theory would then correspond to cutting off this 
iteration at a finite order, which can be justified if $\Omega(t)$ 
changes only very little. 
For the scalar field in Eq.~(\ref{cosmol-eom}) this perturbative 
treatment should be applicable in the ultra-relativistic limit, 
i.e., as long as the mass is much smaller than the wave-number.  

In many cases, however, another approximation -- the WKB method -- 
is more useful. \index{WKB method}
This method can be applied if the rate of change of $\Omega(t)$,  
e.g., the expansion of the universe, is much slower than the internal 
frequency $\Omega(t)$ itself. 
Writing 
\bea\label{cosmol-frequency}
\Omega(t)=\Omega_0 f(\omega t)
\,,
\ea
with some dimensionless function $f$ of order one, the WKB limit corresponds 
to $\Omega_0\gg\omega $. \index{WKB limit}
In terms of the reflection coefficient $R=\beta/\alpha$ mentioned earlier, 
we get 
\bea\label{cosmol-reflection}
\dot R=\frac{\dot\Omega}{2\Omega}
\left(e^{2i\varphi}-R^2e^{-2i\varphi}\right)
\,,
\ea
which is known as Riccati equation. \index{Riccati equation}
Again, this equation is still exact but unfortunately non-linear.
Neglecting the quadratic term $R^2$ would bring us back to perturbation 
theory. 
In the WKB-limit, \index{WKB limit}
the phase factors $e^{\pm2i\varphi}$ are rapidly oscillating 
and the magnitude of $R$ can be estimated by going to the complex plane.
Re-writing the Riccati equation \index{Riccati equation}
(\ref{cosmol-reflection}) as 
\bea\label{cosmol-continuation}
\frac{dR}{d\varphi}=\frac12
\left(e^{2i\varphi}-R^2e^{-2i\varphi}\right)
\frac{d\ln\Omega}{d\varphi}
\,,
\ea
we may use an analytic continuation $\varphi\to\varphi+i\chi$ to see that 
$R$ becomes exponentially suppressed $R\sim e^{-2\chi}$.
How strongly it is suppressed depends on the point where the analytic 
continuation breaks down. 
Since $e^{\pm2i\varphi}$ is analytic everywhere, this will be determined 
by the term $\ln\Omega$.
Typically, the first non-analytic points $t_*$ encountered are the zeros 
of $\Omega$, i.e., where $\Omega(t_*)=0$.
In the case of barrier reflection, these points where $\Omega=0$, i.e., 
where $V=E$, lie on the real axis and correspond to the classical turning 
points in WKB. \index{WKB turning points}
In our case, we have scattering above the barrier and thus these points become 
complex -- but are still analogous to the classical turning points in WKB.
\index{WKB turning points}
Consequently, we find\footnote{In fact, it can be shown that 
Eq.~(\ref{cosmol-exponentially}) becomes exact in the adiabatic limit 
\index{adiabatic limit}
$\omega/\Omega\downarrow0$, i.e., the pre-factor in front of the exponent 
tends to one, see, e.g., [Massar \& Parentani, 1998; Davis \& Pechukas, 1976].}
\bea\label{cosmol-exponentially}
R=\frac{\beta}{\alpha}
\sim e^{-2\chi_*}
=\exp\left\{-2\Im\left(
\int_0^{t_*} dt'\Omega(t')
\right)\right\}
\,.
\ea
If there is more than one turning point, the one with the smallest $\chi_*>0$, 
i.e., closest to the real axis (in the complex $\varphi$-plane) dominates.
If these multiple turning points have similar $\chi_*>0$, there can be 
interference effects between the different contributions, see, e.g., 
[Dumlu \& Dunne, 2010].

\section{Adiabatic expansion and its breakdown}

\index{adiabatic expansion}

\begin{figure}[hbt]
\begin{center}
\epsfig{file=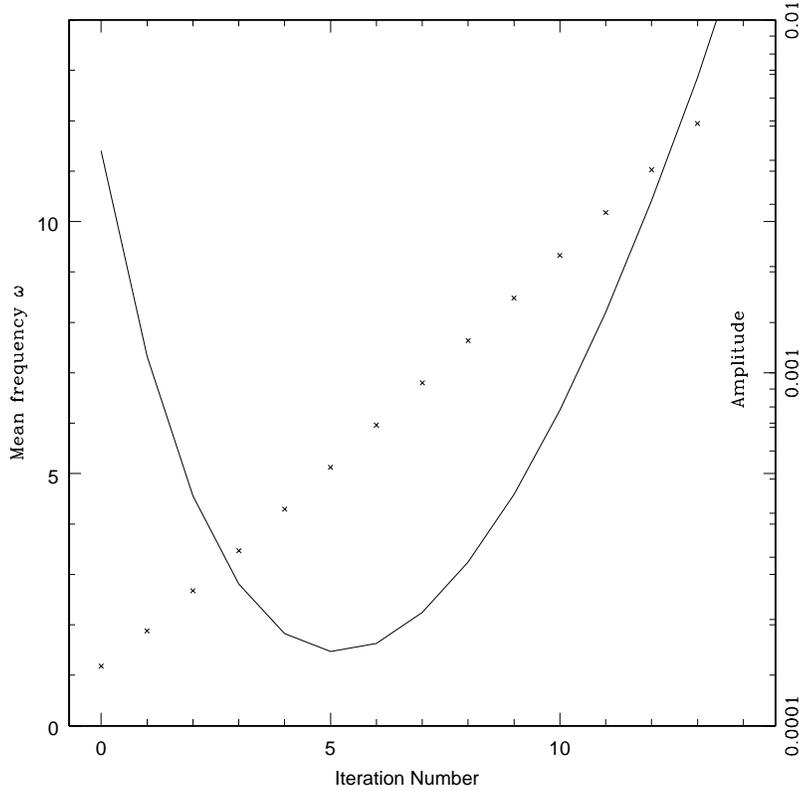,width=.7\textwidth}
\caption{Sketch of the effective external frequencies $\omega_n^{\rm eff}$
(crosses) and amplitudes $\xi_n$ (solid line) depending on the iteration 
number $n$ obtained numerically for a concrete example. 
One can observe that $\omega_n^{\rm eff}$ grows approximately linearly with 
$n$ while $\xi_n$ first decreases but later (for $n>5$) increases again.}
\label{cosmo-picture}
\end{center}
\end{figure}

Note that we could repeat steps (\ref{cosmol-first-order}) till 
(\ref{cosmol-evolution}) and expand the solution $\f{u}(t)$
into the first-order adiabatic eigen-states \index{adiabatic eigen-states}
instead of the instantaneous eigen-vectors $\f{u}_\pm$.
To this end, let us re-write (\ref{cosmol-evolution}) as 
\bea\label{cosmol-re-writing}
\frac{d}{dt}
\left(
\begin{array}{c}
\alpha(t)e^{+i\varphi(t)} \\
\beta(t)e^{-i\varphi(t)}
\end{array}
\right)
=
\f{\dot w}
=
\left(
\begin{array}{cc}
i\Omega & \dot\Omega/(2\Omega) \\
 \dot\Omega/(2\Omega)& - i\Omega
\end{array}
\right)
\cdot
\left(
\begin{array}{c}
\alpha(t)e^{+i\varphi(t)} \\
\beta(t)e^{-i\varphi(t)}
\end{array}
\right)
=
\f{N}\cdot\f{w}
\,.
\ea
The eigen-vectors of the matrix $\f{N}$ are the first-order adiabatic 
eigen-states \index{adiabatic eigen-states}
$\f{w}_\pm$ and the eigen-frequencies 
$\f{N}\cdot\f{w}_\pm=\pm i \Omega_{\rm ad}\f{w}_\pm$ 
are renormalized to 
\bea\label{cosmol-renormalized}
\Omega_{\rm ad}
=
\Omega\sqrt{1-\frac{\dot\Omega^2}{4\Omega^4}}
\,.
\ea
Assuming $\alpha_{\rm in}=1$ and $\beta_{\rm in}=0$, the system stays 
in the adiabatic eigen-state \index{adiabatic eigen-states}
$\f{w}_+$ to lowest order in $\omega/\Omega$ and we get 
\bea\label{cosmol-adiabatic}
\alpha(t)=1+\ord\left(\frac{\omega^2}{\Omega^2}\right)
\,,\;
\beta(t)=-\frac{i}{4}\frac{\dot\Omega}{\Omega^2}
+\ord\left(\frac{\omega^2}{\Omega^2}\right)
\,.
\ea
This adiabatic expansion \index{adiabatic expansion}
into powers of $\omega/\Omega$ can be continued and gives 
terms like $\dot\Omega^2/\Omega^4$ and $\ddot\Omega/\Omega^3$ to the next 
order in $\omega/\Omega$ (see below). 
One should stress that this expansion is {\em not} the same as in 
(\ref{cosmol-iterative}) since it is local -- i.e., only contains 
time-derivatives --  
while (\ref{cosmol-iterative}) is global -- i.e., contains time-integrals.
Since all terms of the adiabatic expansion \index{adiabatic expansion}
(\ref{cosmol-adiabatic}) are local, they cannot describe particle creation 
-- which depends on the whole history of $\Omega(t)$.
In terms of the adiabatic expansion \index{adiabatic expansion}
into powers of $\omega/\Omega$, particle creation is a non-perturbative 
\index{non-perturbative} effect, i.e., it is exponentially 
suppressed, see Eq.~(\ref{cosmol-exponentially})
\bea\label{cosmol-non-perturbativ}
R\sim\exp\left\{-\ord\left(\frac{\Omega}{\omega}\right)\right\}
\,,
\ea
and thus cannot be found be a Taylor expansion into powers of $\omega/\Omega$.
For any finite ratio of $\omega/\Omega$, this also means that the 
adiabatic expansion \index{adiabatic expansion}
(into powers of $\omega/\Omega$) must break down at some point. 
To make this argument more precise, let us re-write 
Eq.~(\ref{cosmol-re-writing}) in yet another form 
\bea\label{cosmol-iteration-start}
\frac{d\f{w}}{dt}
=
\f{N}\cdot\f{w}
=
\Lambda
\left(
\begin{array}{cc}
i\cosh(2\xi) & \sinh(2\xi) \\
\sinh(2\xi) & - i\cosh(2\xi) 
\end{array}
\right)
\cdot\f{w}
\,.
\ea
In this representation, the eigen-values of $\f{N}$ are given by 
$\pm i\Lambda$ and the eigen-vectors read 
\bea\label{cosmol-iteration-eigenvectors}
\f{w}_+
=
\left(
\begin{array}{c}
\cosh\xi \\
-i\sinh\xi
\end{array}
\right)
\,,\quad
\f{w}_-
=
\left(
\begin{array}{c}
\sinh\xi \\
-i\cosh\xi
\end{array}
\right)
\,.
\ea
Decomposing the solution $\f{w}(t)$ into these eigen-vectors 
\bea\label{cosmol-iteration-decompose}
\f{w}(t)=a(t)\f{w}_+(t)+b(t)\f{w}_-(t)
\,,
\ea
and using $\f{\dot w}_+=\dot\xi\f{w}_-$ as well as 
$\f{\dot w}_-=\dot\xi\f{w}_+$, we find 
\bea\label{cosmol-iteration-end}
\frac{d}{dt}
\left(
\begin{array}{c}
a \\
b
\end{array}
\right)
=
\left(
\begin{array}{cc}
i\Lambda & -\dot\xi \\
-\dot\xi & - i\Lambda
\end{array}
\right)
\cdot
\left(
\begin{array}{c}
a \\
b 
\end{array}
\right)
\,.
\ea
This is the same form as Eq.~(\ref{cosmol-iteration-start}) if we change 
$\Lambda$ and $\xi$ accordingly.
Thus, by repeating this procedure, we get the iteration law 
\bea\label{cosmol-iteration-law}
\Lambda_{n+1}=\sqrt{\Lambda_{n}^2-\dot\xi_n^2} 
\,,\quad
\xi_{n+1}=-\frac12\,
{\rm arctanh}\left(\frac{\dot\xi_n}{\Lambda_n}\right)
\,.
\ea
By this iteration, we go higher and higher up in the adiabatic expansion
\index{adiabatic expansion}
since $\xi_n$ always acquires an additional factor of $\omega/\Omega$.
Thus, for $\omega\ll\Omega$, the values of $\xi_n$ quickly decay with a 
power-law $\xi_n=\ord([\omega/\Omega]^n)$ initially.
As we go up in this expansion, however, the effective rate of change of 
$\xi_n$ increases.
For example, if $\Omega(t)$ has one global maximum (or minimum) and otherwise 
no structure, the time-derivative $\dot\Omega/(2\Omega^2)=\tanh(2\xi_1)$ 
has two extremal points and a zero in between.
By taking higher and higher time derivatives, more and more extremal points 
and a zeros arise and thus the effective frequency $\omega_n^{\rm eff}$
of $\xi_n(t)$ increases roughly linearly with the number $n$ of iterations 
$\omega_n^{\rm eff}=\ord(n\omega)$.
Furthermore, the adiabatically \index{adiabatic eigen-states}
renormalized eigen-values $\Lambda_n$ decrease with each iteration. 
Thus, after approximately $n=\ord(\Omega/\omega)$ iterations, 
the effective frequency $\omega_n^{\rm eff}$ becomes comparable to 
the internal frequency $\Lambda_n$.
At that point, the adiabatic expansion \index{adiabatic expansion}
starts to break down.
Estimating the order of magnitude of $\xi_n$ at that order gives
\bea\label{cosmol-xi-breakdown}
\xi_n
=
\ord\left(\left[\frac{\omega}{\Omega}\right]^n\right)
=
\ord\left(\left[\frac{\omega}{\Omega}\right]^{\ord(\Omega/\omega)}\right)
\,.
\ea
Since the effective external $\omega_n^{\rm eff}$ and internal $\Lambda_n$ 
frequencies are comparable and $\xi_n$ is very small, we may just use 
perturbation theory to estimate $\beta$ and we get $\beta=\ord(\xi_n)$, 
i.e., the same exponential suppression as in 
Eq.~(\ref{cosmol-non-perturbativ}). 
If we would continue the iteration beyond that order, the $\xi_n$ would start 
to increase again -- which the usual situation in an asymptotic expansion,
see Figure~\ref{cosmo-picture}. \index{asymptotic expansion}
Carrying on the iteration too far beyond this point, the 
$\dot\xi_n^2$ exceed the $\Lambda_n^2$ and thus we have barrier 
penetration instead of propagation over the barrier 
(as occurs for all orders below this value of $n$). 
In this procedure, it is this barrier penetration which gives the mixing 
of positive and negative pseudo-norm, and the creation of particles. 
Were the system to remain as propagation over the barrier for all
orders $n$ in this adiabatic expansion, one would have no particle creation.

\section{Example: inflation}\label{cosmo-inflation}

\index{inflation}

As an illustrative example, let us consider a minimally coupled massive scalar 
field in 3+1 dimensions -- which could be the inflaton field 
\index{inflaton field}
(according to our standard model of cosmology).
Again, we start with the Friedmann-Robertson-Walker metric (\ref{cosmol-frw})
\index{Friedmann-Robertson-Walker metric}
with a scale factor $a(\tau)$ and obtain the equation of motion 
\bea\label{cosmol-3+1}
\left(
\frac{1}{a^3(\tau)}\,
\frac{\partial}{\partial\tau}\,
a^3(\tau)\,
\frac{\partial}{\partial\tau}
-
\frac{1}{a^2(\tau)}\,
\na^2
+m^2
\right)\Phi=0
\,.
\ea
Rescaling the field $\phi(\tau,\f{r})=\mho(\tau)\Phi(\tau,\f{r})$
with $\mho(\tau)=a^{3/2}(\tau)$ and applying a spatial Fourier transform, 
we obtain the same form as in Eq.~(\ref{cosmol-oscillator}) 
\bea\label{cosmol-rescaling}
\left(
\frac{d^2}{d\tau^2}
+
\frac{\f{k}^2}{a^2(\tau)}
+m^2
-\frac{1}{\mho(\tau)}\,\frac{d^2\mho(\tau)}{d\tau^2}
\right)\phi_k=0
\,.
\ea
In the standard scenario of inflation,\index{inflation}
the space-time can be described by the de~Sitter metric 
\index{de~Sitter metric}
$a(\tau)=\exp\{H\tau\}$ to a very good approximation,
where $H$ is the Hubble parameter. \index{Hubble parameter}
In this case, the effective potential $\ddot\mho/\mho$ just becomes a 
constant $(3H/2)^2$ and the frequency $\Omega(\tau)$ reads
\bea\label{cosmol-frequency-3+1}
\Omega^2(\tau)=\frac{\f{k}^2}{a^2(\tau)}+m^2-\frac{9H^2}{4}
\,.
\ea
Inserting $a(\tau)=\exp\{H\tau\}$, we see that modes with different
$k$-values follow the same evolution -- just translated in time.  
(This fact is related to the scale invariance of the created $k$ spectrum.)  
Initially, this frequency is dominated by the $\f{k}^2$ term and we have 
$\dot\Omega/\Omega=-H$ which means that we are in the WKB regime 
\index{WKB limit} 
$\dot\Omega/\Omega\ll\Omega$.
However, due to the cosmological red-shift, this $\f{k}^2$ term decreases 
with time until the other terms become relevant.
Then the behavior of the modes depends on the ratio $m/H$.
For $m\gg H$, the modes remain adiabatic \index{adiabatic limit}
(i.e., stay in the WKB regime) \index{WKB limit} 
and thus particle creation is exponentially suppressed.
If $m$ and $H$ are not very different, but still $m>3H/2$ holds, 
the modes are adiabatic \index{adiabatic limit}
again for large times -- but for intermediated times, 
the WKB expansion \index{WKB method} 
breaks down, leading to a moderate particle creation.
For $m<3H/2$, on the other hand -- which is (or was) supposed to be the 
case during inflation \index{inflation}
-- the frequency $\Omega(\tau)$ goes to zero at 
some time and becomes imaginary afterwards.
This means that we get a barrier penetration (tunneling) problem 
where the modes $\phi_k(\tau)$ do not oscillate but evolve exponentially 
in time $\phi_k(\tau)\propto\exp\{\pm\tau\sqrt{9H^2/4-m^2}\}$.
Here one should remember that the original field does not grow
exponentially due to the re-scaling with the additional factor 
$\mho(\tau)=a^{3/2}(\tau)$.
This behavior persists until the barrier vanishes, i.e., the expansion 
slows down (at the end of the inflationary period) \index{inflation}
and thus the effective potential $\ddot\mho/\mho$ drops below the mass term.
After that, the modes start oscillating again.  
However, in view of the barrier penetration (tunneling) over a 
relatively long time (distance), we get reflection coefficients $R$
which are not small but extremely close to unity $R\approx1$.
This means that the Bogoliubov coefficients $\alpha$ and $\beta$ 
\index{Bogoliubov coefficients} 
are huge -- i.e., that we have created a tremendous amount of 
particles out of the initial vacuum fluctuations.
According to our understanding, precisely this effect is responsible 
for the creation of the seeds for all structures in our Universe.
Perhaps the most direct signatures of this effect are still visible 
today in the anisotropies of the cosmic microwave background radiation. 
\index{cosmic microwave background}

An alternative picture of the mode evolution in terms of a damped 
harmonic oscillator can be obtained from 
the original field in Eq.~(\ref{cosmol-3+1})
\bea\label{cosmol-damped}
\left(
\frac{d^2}{d\tau^2}
+3H\,\frac{d}{d\tau}
+e^{-2H\tau}\f{k}^2
+m^2
\right)\Phi_k=0
\,.
\ea
Initially, the term $e^{-2H\tau}\f{k}^2$ dominates and the modes 
oscillate.
Assuming $m\ll H$ (which is related to the slow-roll condition of inflation),  
\index{inflation}
the damping term dominates for late times and we get a strongly 
over-damped oscillator, whose dynamics is basically frozen 
(like a pendulum in a very sticky liquid).
The transition happens when $H\sim ke^{-H\tau}$, i.e., when the 
physical wavelength $\lambda=2\pi e^{H\tau}/k$ exceeds the de~Sitter 
\index{de~Sitter metric}
horizon $\propto1/H$ due to the cosmological expansion $e^{H\tau}$. 
After that, crest and trough of a wave lose causal contact and 
cannot exchange energy any more -- 
that's why the oscillations effectively stops. 

As a final remark, we stress that this enormous particle creation 
effect is facilitated by the rapid (here: exponential) expansion and 
the resulting stretching of wavelengths over many many orders of magnitude
(i.e., the extremely large red-shift).
Therefore, a final mode with a moderate wavelength originated from 
waves with extremely short wavelengths initially.
Formally, these initial wavelengths could be easily far shorter 
than the Planck length.
However, on these scales one would expect deviations from the 
theory of quantum fields in classical space-times we used to 
derive these effects.
On the other hand, this problem is not only negative -- it might 
open up the possibility to actually see signatures of new (Planckian) 
physics in high-precision measurements of the cosmic microwave 
background radiation, for example. \index{cosmic microwave background}

\section{Laboratory analogues}\label{Laboratory analogues}

\index{laboratory analogues}

Apart from the observation evidence in the anisotropies of the cosmic 
microwave background radiation \index{cosmic microwave background} 
mentioned above, one may study the phenomenon 
of cosmological particle creation \index{cosmological particle creation}
experimentally by means of suitable laboratory analogues, see, e.g., 
[Unruh, 1981; Barcel\'o, Liberati, \& Visser, 2011]. 
\index{laboratory analogues}
The are two major possibilities to mimic the expansion or contraction of 
the Universe -- a medium at rest with time-dependent properties 
(such as the propagation speed of the quasi-particles) or an expanding medium.
Let us start with the former option and consider linearized and scalar 
quasi-particles (e.g., sound waves) with low energies and momenta propagating 
in a spatially homogeneous and isotropic medium.
Under these conditions, their dynamics is governed by the low-energy 
effective action \index{low-energy effective action}
\bea
\label{cosmo-Leff}
{\cal L}_{\rm eff}
=
\frac12\left(a^2(t)\dot\phi^2+b^2(t)\phi^2+c^2(t)[\na\phi]^2\right)
+\ord(\phi^3)+\ord(\partial^3)
\,.
\ea
Here we assume positive $a^2$ and non-negative $b^2$ and $c^2$ for stability.
The factor $a^2(t)$ can be eliminated by suitable re-scaling of the time 
co-ordinate. 
Then, after a spatial Fourier transform, we obtain the same form as in 
Eq.~(\ref{cosmol-oscillator}).
The quasi-particle excitations $\phi$ in such a medium behave in the same way 
as a scalar field in an expanding or contracting Universe with a possibly
time-dependent potential (mass) term $\propto b^2(t)\phi^2$.
In order to avoid this additional time-dependence of the potential (mass) 
term, the factors $b$ and $c$ must obey special conditions.
For example, Goldstone modes with $b=0$ correspond to a massless scalar field 
in 3+1 dimensions -- whereas the case of constant $c$ is analogous to a 
massive scalar field in 1+1 dimensions. 

As one would intuitively expect, the expansion or contraction of the Universe 
can also be mimicked by an expanding or contracting medium. 
Due to local Galilee invariance, such a medium can also be effectively 
spatially homogeneous and isotropic as in Eq.~(\ref{cosmo-Leff}) 
when described in terms of co-moving co-ordinates. 
For a quite detailed list of references, see 
[Barcel\'o, Liberati, \& Visser, 2011].

There are basically three major experimental challenges for observing 
the analogue of cosmological particle creation 
\index{cosmological particle creation} 
in the laboratory. 
\index{laboratory analogues}
First, the initial temperature should be low enough such that the particles
are produced due to quantum rather than thermal fluctuations. 
Second, one must be able to generate a time-dependence 
(e.g., expansion of the medium) during which the effective action in 
Eq.~(\ref{cosmo-Leff}) remains valid (in some sense) but which is 
also sufficiently rapid to create particles. 
Third, one must be able to detect the created particles and to distinguish
them from the radiation stemming from other sources. 
For trapped ions, \index{trapped ions}
for example (see, e.g., [Sch\"utzhold {\em et al}, 2007]),  
the first and third point (i.e., cooling and detection) is experimental 
state of the art, while a sufficiently rapid but still controlled 
expansion/contraction of the ion trap presents difficulties. 
For Bose-Einstein condensates \index{Bose-Einstein condensates}
(see, e.g., [Barcel\'o, Liberati, \& Visser, 2011] and references therein), 
on the other hand, the first and third points are the main obstacles. 

\newpage

\subsection*{Acknowledgments}

The authors benefited from fruitful discussions, especially with R. Parentani,
during the {\em SIGRAV Graduate School in Contemporary Relativity and 
Gravitational Physics}, IX Edition ``Analogue Gravity'', at the Centro di 
Cultura Scientifica ``A. Volta'', Villa Olmo, in Como (Italy, 2011). 
R.S.\ acknowledges support from DFG 
and the kind hospitality during a visit at the University of British 
Columbia where part of this research was carried out. 
W.G.U.~thanks the Natural Sciences and Engineering Research Council of Canada 
and  the Canadian Institute for Advanced Research, for research support, 
and the University of Duisburg-Essen for their hospitality while part of 
this research was carried out.

\newpage

\section*{References}

\begin{enumerate}

\item
L.C.~Baird, 
{\em New Integral Formulation of the Schr\"odinger Equation}, 
J.\ Math.\ Phys.\ {\bf 11}, 2235 (1970). 

\item
C.~Barcel\'o, S.~Liberati, and M.~Visser,
{\em Analogue Gravity}, 
Living Rev.\ Relativity {\bf 14}, 3 (2011). 

\item
N.~D.~Birrell and P.~C.~W.~Davies,
{\em Quantum Fields in Curved Space},
(Cambridge University Press, Cambridge, England 1982).

\item
J.~P.~Davis and P.~Pechukas,
{\em Nonadiabatic transitions induced by a time?dependent 
Hamiltonian in the semiclassical/adiabatic limit: 
The two?state case}, 
J.\ Chem.\ Phys.\ {\bf 64}, 3129 (1976). 

\item
C.~K.~Dumlu and G.~V.~Dunne, 
{\em Stokes Phenomenon and Schwinger Vacuum Pair 
Production in Time-Dependent Laser Pulses}, 
Phys.\ Rev.\ Lett.\ {\bf 104}, 250402 (2010). 


\item
S.~A.~Fulling,
{\em Aspects of Quantum Field Theory in Curved Space-Time},
(Cambridge University Press, Cambridge, England 1989). 

\item
S.~Massar and R.~Parentani,
{\em Particle creation and non-adiabatic transitions in quantum cosmology},  
Nucl.\ Phys.\ B {\bf 513}, 375 (1998). 

\item
L.~Parker,
{\em Particle creation in expanding universes}, 
Phys.\ Rev.\ Lett.\ {\bf 21}, 562 (1968).

\item
E.~Schr\"odinger, 
{\em The proper vibrations of the expanding Universe}, 
Physica {\bf 6}, 899 (1939). 

\item
R.~Sch\"utzhold, M.~Uhlmann, L.~Petersen, H.~Schmitz, A.~Friedenauer,
and T.~Sch\"atz, 
{\em Analogue of cosmological particle creation in an ion trap}, 
Phys.\ Rev.\ Lett.\ {\bf 99}, 201301 (2007). 

\item
W.~G.~Unruh,
{\em Experimental Black Hole Evaporation?},
Phys.\ Rev.\ Lett.\ {\bf 46}, 1351 (1981).

\item
R.~M.~Wald, 
{\em Quantum Field Theory in Curved Spacetime and Black Hole Thermodynamics}, 
(University of Chicago Press, Chicago, 1994).

\end{enumerate}

\newpage

\printindex

\end{document}